# The Integrated Impact Indicator (*I3*) Revisited:

# A Non-Parametric Alternative to the Journal Impact Factor

Loet Leydesdorff,*[a] Lutz Bornmann,[b] and Jonathan Adams[c]

**Abstract**

We propose the *I3\** indicator as a non-parametric alternative to the Journal Impact Factor (JIF) and *h*-index. We apply *I3\** to more than 10,000 journals. The results can be compared with other journal metrics. *I3\** is a promising variant within the general scheme of non-parametric indicators *I3* introduced previously: it provides a single metric which correlates with both impact in terms of citations (*c*) and output in terms of publications (*p*). We argue for weighting using four percentile classes: the top-1% and top-10% as excellence indicators; the top-50% and bottom-50% as output indicators. Like the *h*-index, which also incorporates both *c* and *p*, *I3\**-values are size-dependent; however, division of *I3\** by the number of publications (*I3\*/N*) provides a size-independent indicator which correlates strongly with the two- and five-year Journal Impact Factors (JIF2 and JIF5). Unlike the *h*-index, *I3\** correlates significantly with *both* the total number of citations and publications. The values of *I3\** and *I3\*/N* can be statistically tested against the expectation or against one another using chi-square tests or effect sizes. A template (in Excel) is provided online for relevant tests.

**Keywords**: journal indicator, percentile, citation analysis, *I3\**, Journal Impact Factor

[a] Amsterdam School of Communication Research, University of Amsterdam, PO Box 15793, 1001 NG Amsterdam, The Netherlands; l.a.leydesdorff@uva.nl
[b] Max Planck Society, Administrative Headquarters, Hofgartenstr. 8, 80539 Munich, Germany; bornmann@gv.mpg.de
[c] The Policy Institute at King's, King's College London, 22 Kingsway, London, WOS CATEGORY2B 6LE, UK; Institute for Scientific Information, Clarivate Analytics, Blackfriars Road, London, UK; jonathan.adams@kcl.ac.uk.

# 1. Introduction

Citations create links between publications; but to relate citations to publications as two different things, one needs a model (for example, an equation). The Journal Impact Factor (JIF) indexes only one aspect of this relationship: citation impact. Using the *h*-index, papers with at least *h* citations are counted. One can also count papers with $h^2$ or $h/2$ citations (Egghe, 2008). This paper is based on a different and, in our opinion, more informative model: the Integrated Impact Indicator *I3*.

The two-year JIF was outlined by Garfield & Sher (1963; cf. Garfield, 1955; Sher & Garfield, 1965) at the time of establishing the Institute for Scientific Information (ISI). JIF2 is defined as the number of citations in the current year (*t*) to any of a journal's publications of the two previous years (*t*-1 and *t*-2), divided by the number of citable items (substantive articles, reviews, and proceedings) in the same journal in these two previous years. Although not strictly a mathematical average, JIF2 provides a functional approximation of the mean early citation rate per citable item. A JIF2 of 2.5 implies that, *on average*, the citable items published one or two years ago were cited two and a half times. Other JIF variants are also available; for example, JIF5 covers a five-year window.[1]

The central problem that led Garfield (1972; 1979) to use the JIF when developing the *Science Citation Index,* was the selection of journals for inclusion in this database. He argued that citation analysis provides an excellent source of information for evaluating journals. The choice of a two-year time window was based on experiments with the Genetics Citation Index and the early Science Citation Index (Garfield, 2003, at p. 364; Martyn & Gilchrist, 1968). However, one possible disadvantage of the short term (two years) could be that "the journal impact factors enter the picture when an individual's most recent papers have not yet had time to be cited" (Garfield, 2003, p. 365; cf. Archambault & Larivière, 2009). Bio-medical fields have a fast-moving research front with a short citation cycle, and JIF2 may be an appropriate measure for such fields but less so for other fields (Price, 1970). In the 2007 edition of Journal Citation Reports (reissued for this reason in 2009) a five-year JIF (JIF5, considering five instead of only two publication years) was added to balance the focus on short-term citations provided by JIF2 (Jacsó, 2009; cf. Frandsen & Rousseau, 2005).[2]

---

[1] A journal that publishes many items that do not report substantive research, but nonetheless attract citations, can inflate its JIF (Moed & van Leeuwen, 1996).

The skew in citation distributions provides another challenge to the evaluation (Seglen, 1992; 1997). The mean of a skewed distribution provides less information than the median as a measure of central tendency. To address this problem, McAllister, Narin, & Corrigan (1983, at p. 207) proposed the use of percentiles or percentile classes as a non-parametric indicator (Narin, 1987;[3] see later: Bornmann & Mutz, 2011; Tijssen, Visser, & van Leeuwen, 2002). Using this non-parametric approach, and on the basis of a list of criteria provided by Leydesdorff, Bornmann, Mutz, & Opthof (2011), two of us first developed the Integrated Impact Indicator (*I3*) based on the integration of the quantile values attributed to each element in a distribution (Leydesdorff & Bornmann, 2011).

---

[2] On the basis of questionnaires among faculty, Bensman (2007) concluded that total cites would be a better indicator of the longer-term influence of a journal (Gross & Gross, 1927).
[3] Narin (1987) normalized on the basis of a scheme developed by him (Narin, 1976).

Since *I3* is based on integration, the development of *I3* presents citation analysts with a construct fundamentally different from a methodology based on averages. An analogy that demonstrates the difference between integration and averaging is given by basic mechanics: the impact of two colliding bodies is determined by their combined mass and velocity, and not by the average of their velocities. So, it can be argued that the gross impact of the journal as an entity is the combined volume and citation of its contents (articles and other items); but not an average. Journals differ both in size (the number of published items) and in the skew and kurtosis of the distribution of citations across items. A useful and informative indicator for the comparison of journal influences should respond to these differences. A citation average cannot reflect the variation in both publications and citations but an indicator based on integration can do so.

One route to indexing both performance and impact via a single number has been provided by the *h*-index (Hirsch, 2005) and its variants (e.g., Bornmann *et al*., 2011; Egghe, 2008). However, the *h*-index has many drawbacks, not least mathematical inconsistency (Marchant, 2009; Waltman & Van Eck, 2012). Furthermore, Bornmann, Mutz, & Daniel (2008) showed that the *h*-index is mainly determined by the number of papers (and not by citation impact). In other words, the impact dimension of a publication set may not be properly measured using the *h*-index. One aspect that *I3* has in common with the *h*-index is that the focus is no longer on impact as an attribute but on the information production process (Egghe & Rousseau, 1990; Ye *et al*., 2017). This approach could be applied not only to journals but also to other sets of documents with citations such as the research portfolios of departments or universities. In this study, however, we focus on journal indicators.

At the time of our previous paper about *I3* (Leydesdorff & Bornmann, 2011), we were unable to demonstrate the generic value of the non-parametric approach because of limited data access. Recently, however, the complete *Web of Science* became accessible under license to the Max Planck Society (Germany). This enables us to compare *I3*-values across the database with other journal indicators such as JIF2 and JIF5, total citations (*NCit*), and numbers of publications (*NPub*). The choice for journals as units of analysis provides us with a rich and well-studied domain.

Our approach based on percentiles can be considered as the development of "second generation indicators" for two reasons. First, we build on the first-generation approach that Garfield (1979; 2003; 2006) developed for the selection of journals. Second, the original objective of journal selection is very different from the purposes of research evaluation to which JIF has erroneously ben applied (e.g., Alberts, 2013). The relevant indicators should accordingly be appropriately sophisticated.

**The weighting scheme**

In this study, we introduce *I3\**—a variant within the general *I3* scheme—by proposing a weighting scheme of percentile classes. We elaborate on Bornmann & Mutz (2011) who counted six percentile classes with weights from one to six. Since that publication, however, several threads of work have clarified the position of the top-10% and top-1% categories as proxies for excellence. On the basis of this literature (e.g., Bornmann, 2014), our basic assertion is that a paper in the top-1% class can be weighted at ten times the value of a paper in the top-10% class.

It follows log-linearly that a top-1% paper weighs 100 times more than a paper at the bottom. This weighting scheme reflects the highly-skewed nature of citation distributions. We add, as a second assertion, a weighting to distinguish between papers in the top-50% (weight = 2) and bottom-50% (weight = 1). The dividing line between bottom-50% and top-50% is less pronounced than the line between an averagely-cited paper and an exceptionally-cited one.

Figure 1 and Table 1 clarify the correspondence between the approaches. (We will show the differences empirically in a later section.) In Figure 1 the left axis is logarithmic—that is, log(1) to log(100) —whereas the right axis is linear (one to six). In the original scheme of Mutz & Bornmann (2011), the relative weighting of a top-1% and top-10% paper was only 6 : 4.5 (equivalent to 4:3) whereas we apply 10 : 1 (= 10) in the new scheme. Using quantiles (Leydesdorff & Bornmann, 2011), the relation between a top-1% and top-10% paper would only be 99 : 89 (= 1.1).

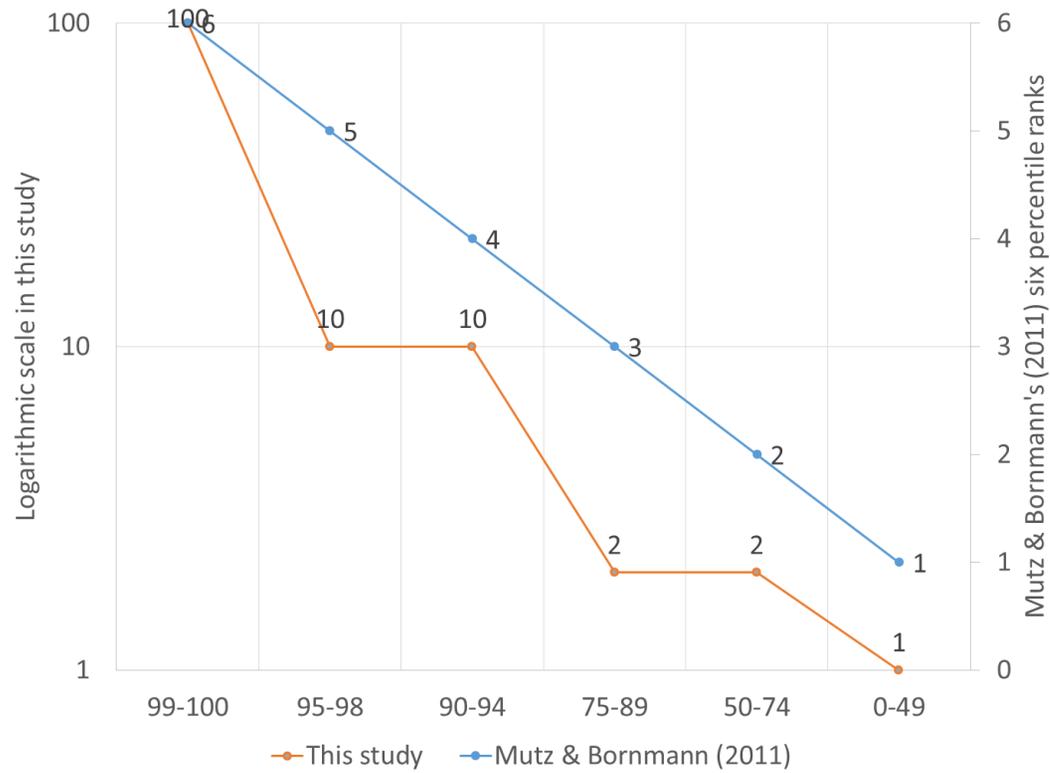

**Figure 1**: Weighting factors of the percentile ranks in Mutz & Bornmann (2011) and this study.

**Table 1**: Weighting factors of the percentile ranks in Mutz & Bornmann (2011) and this study.

| Percentile ranks | Mutz & Bornmann (2011) | This study |
|---|---|---|
| 99-100 | 6 | 100 |
| 95-98 | 5 | 10 |
| 90-94 | 4 | 10 |
| 75-89 | 3 | 2 |
| 50-74 | 2 | 2 |
| 0-49 | 1 | 1 |

In other words, we distinguish between *I3* as a general scheme and a possible family of specific weighting schemes. The latter are applications for specific evaluation contexts. In general, *I3* can be written as follows:

$$I3(PR_1\text{-}W_1, PR_2\text{-}W_2, \ldots PR_n\text{-}W_n)$$

where *PR* defines the lower threshold of the respective percentile rank class and *W* the corresponding weight; *n* is the number of classes and weights, respectively. In this notation, the scheme proposed by Bornmann & Mutz (2011)—at the time called *PR6*—can be written as follows: *I3*(99-6, 95-5, 90-4, 75-3, 50-2, 0-1); and the scheme in this paper (*I3\**) can be formalized as *I3*(99-100, 90-10, 50-2, 0-1). However, the scheme can be used more broadly for percentile-based indicators: the top-10% so-called excellence indicator (e.g., Bornmann *et al.*, 2012; Waltman *et al.*, 2012), for example, can be formalized as the special case *I3*(90-1). In this study, we propose a new variant which we denote as *I3\**; *I3* can be considered as a pragmatic shorthand of *I3*(99-100, 90-10, 50-2, 0-1).

As is the case for all *I3* evaluations, *I3\** is size-dependent: it scales *ceteris paribus* with journal size. By dividing *I3\** by the number of elements of the distribution $N = \Sigma_i \, n_i$ (of documents), a size-independent equivalent can be generated. Not surprisingly, this latter measure is highly correlated with JIF2 and JIF5. In other words, $I3^*_i / N_i$ provides the journal-specific expected *I3\** value of a paper published in journal *i*. This value can be used as a benchmark for testing whether the observed citation count for a specific paper is above or below expectation.

Note that we test expected citation rates against observed ones at the level of a sample (e.g., a journal). Consequently, our approach avoids the "ecological fallacy" of using a journal characteristic as an expected value to compare with observed values derived from the individual papers published in the respective journal (Robinson, 1950; Kreft & de Leeuw, 1988; cf. Waltman & Traag, 2017). The observed values are not estimated on the basis of a journal characteristic, but are measured in order to inform the expectation.

**2. Methods**

2.1. *Data*

Data were harvested at the Max Planck Digital Library (MPDL) in-house database of the Max Planck Society during the period October 15-29, 2018. This database contains an analytically enriched copy of the *Sciences Citation Index*-Expanded (SCI-E), the *Social Sciences Citation Index* (SSCI), and the *Arts and Humanities Citation Index* (AHCI). Citation count data can be normalized for the Clarivate *Web of Science* Subject Categories (WoS Categories) and theoretically could be based on whole-number counting or fractional counting in the case of more than a single co-author. The unit of analysis in this study, however, is the individual paper to which citation counts are attributed irrespective of whether the paper is single- or multi-authored.

The citation window in the in-house database was the period to the end of 2017, at the time of the data collection. We collected substantive items (articles and reviews) using the publication year 2014 with a three-year citation window to the end of 2017. The results were checked against a similar download for the publication year 2009, that is, five years earlier. The year 2014 was chosen as the last year with a complete three-year citation window at the time of this research (October-November, 2018); the year 2009 was chosen because it is the first year after the update of WoS to its current version 5.

2.1.1. Non-normalized data

The in-house database contains many more journals than the *Journal Citation Reports* (which form the basis for the computation of JIF). In order to be able to compare between *I3\**-values and other indicators, we use only the subset of publications in the 11,761 journals contained in the *Journal Citation Reports* (JCR) 2014. These journals all have JIFs and other standard indicators. Of these journals, 11,149 are unique in the SCI-E and SSCI, and the overlap between SSCI and SCI-E is 612 journals. Another 207 journals could not be matched unequivocally on the basis of journal name abbreviations in the in-house database and JCR, so that our sample is 10,942 journals. Note that we are using individual-journal attributes so that the inclusion or exclusion of a specific journal does not affect the values for the other journals under study.

We collected the data as follows. On the basis of the number of papers (articles and reviews, excluding non-academic ephemera such as editorials) in a specific year (in this case: 2014), we identified the threshold number of citations at category boundaries, e.g. the lower boundary of the 1% most-frequently cited papers. If there are, for example, a total of 100,000 papers in a year, then one thousand of them should belong to the most-cited-1% for obvious stochastic reasons. If the papers are ranked by descending citation count, then the citation count of the 1,000th paper is the threshold value (Ahlgren *et al.*, 2014). For each journal, the number of papers in this set can be counted. By counting the number of papers with a citation count <u>exceeding</u> this threshold value, the problem of ties is circumvented. However, there is a possibility that more than 1,000 papers may thereby be included in the top-1% because there are several papers with the same value as the threshold (in 2014, e.g., 1.03% of the papers instead of exactly 1%). The same applies to the other top-*x*% classes.

In summary, we harvest the top-1%, top-10%, top-50%, and bottom-50% publication scores for each journal by first determining the thresholds of these percentile classes for the entire database and, second, by counting each journal's participation in the respective layers of the database. Using a dedicated routine, the data are organized in a relational database with JCR-2014 data. The tables resulting from the analyses can be read into standard software (e.g., Excel, SPSS) for further processing and statistical analysis.

2.1.2 Normalized data

Citation counts are also field-normalized in the in-house database using the WoS Categories, because citation rates differ between fields. These field-normalized scores are available at individual document level for all publications since 1980. The *I3\** indicator calculated with field-normalized data will be denoted as *I3\*F*—pragmatically abbreviating *I3\*F*(99-100, 90-10, 50-2, 0-1) in this case. Some journals are assigned to more than a single WOS CATEGORY: in these instances, the journal items and their citation counts are fractionally attributed. In the case of ties at the thresholds of a top-*x*% class of papers (see above), the field-normalized indicators have been calculated following Waltman & Schreiber (2013). Thus, the in-house database shows whether a paper belongs to the top-1%, top-10%, or top-50% of papers in the corresponding WoS Categories. Papers at the threshold separating the top from the bottom are fractionally assigned to the top paper set.

*2.2. Statistics*

Table 2 shows how to calculate $I3^*$ based on publication numbers using *PLOS One* as an example. The publication numbers in the first columns (*a* and *b*) are obtained from the in-house database of the Max Planck Society. These are the numbers of papers in the different top-$x$%-classes. Since the publication numbers in the higher classes are subsets of the numbers in the lower classes, the percentile classes are corrected (by subtraction) to avoid double counting. The resulting values in each distinct class are provided in the columns *c* and *d*. The distinct class counts are then multiplied by the appropriate weights. In the last step of calculating $I3^*$, the weighted numbers of papers in the distinct classes are summed into $I3^*$. In this case, $I3^* = 78,733$ (non-normalized) and $I3^*F = 53,570.256$ (field-normalized).

**Table 2**: *PLOS One* data as an example of the calculation of *I3\**, based on non-normalized and field-normalized values

|  | Data from the in-house database | | Distinct classes | Number of papers in distinct classes | | Weights | I3* and I3*F | |
| --- | --- | --- | --- | --- | --- | --- | --- | --- |
|  | Non-normalized (a) | Field-normalized (b) | Percentile rank classes | Non-normalized (c) | Field-normalized (d) |  | Non-normalized (f) | Field-normalized (g) |
| **Top 1%** | 91 | 14.000 | 99-100 | 91 | 14.000 | x 100 = | 9100 | 1400 |
| **Top 10%** | 2,545 | 926.821 | 90-98 | 2,454 | 912.821 | x 10 = | 24,500 | 9,128.21 |
| **Top 50%** | 20,141 | 14,853.688 | 50-89 | 17,506 | 13,926.867 | x 2 = | 35,192 | 27,853.73 |
|  |  |  | 0-49 | 9,901 | 15,188.312 | x 1 = | 9,901 | 15,188.31 |
| **Total** | 30,042 | 30,042 |  | 30,042 | 30,042 |  | **78,733** | **53,570.26** |

The maximal *I3\** is ((30,042 * 100) + (0 * 10) + (0 * 2) + (0 * 1) = ) 3,004,200 whereby all papers in the journal would belong to the 1% most frequently cited papers in the corresponding fields. With *I\** = 53,570.256, the journal reaches 1.78% of this maximum. Without field-normalization, this is 2.62%. In other words, there is ample room for improvement.[7]

As noted, *I3\** can be divided by *N*, the number of publications (which is by definition equal to the sum of the numbers in the four percentile classes). *I3\*/N* is based on relative frequencies, since the number in each term ($n_i$) is divided by *N* (= $\Sigma_i n_i$ ). One can expect *I3\*/N* to no longer be size-dependent and thus to have applications different from *I3\**, as we shall show below. We focus on *I3\** in this paper; we will discuss potential applications of *I3\*/N* in a later paper.

---

[7] Analogously, the minimal *I3\** which *PLOS One* 2014 could reach is 30,042; all publications would in this case belong to the bottom-50% papers and thus be weighted only with a one (0 * 100 + 0 * 10 + 0 * 2 + 30,042 * 1 = 30,042).

We have applied Spearman rank-correlation analysis and factor analysis (Principal Component Analysis with varimax rotation) to the following variables:[1]

1. total numbers of publications (NPub);
2. citations (NCit);
3. JIF2;
4. JIF5;
5. Non-normalized *I3\**-values (*I3\**);
6. Field-normalized *I3\**-values (*I3\*F*);
7. *I3\*/N* for the non-normalized case (*I3\*/N*).

The results are shown as factor-plots using the first two components as *x*- and *y*-axes. This representation in a two-dimensional map provides a ready means of assessing the results visually.

We chose two components in accordance with our design, but the number of eigenvectors with a value larger than one is also two. The results indicate that the two first eigenvectors explain about 85-90% of the variance in the subsequent analyses. Since the distributions are non-normal, Spearman's rank-order correlations are preferable to Pearson correlations.[2] Note that the factor-analysis is based on Pearson correlations and the results are consequently, in this respect, approximations. Rotated factor matrices and the percentages of explained variance are also provided for each analysis.

---

[1] We checked also for oblique rotation, but the results are very similar.
[2] A non-parametric alternative would be to use multidimensional scaling (MDS, Schiffman, Reynolds, & Young, 1981).

## 3. Results

*3.1 Full set* (journal count, $n = 10{,}942$)

Figure 2 shows the two-dimensional factor plot of the data provided numerically in Table 3. The first two factors explain 87.5% of the variance. The correlation between *I3\** and its field-normalized equivalent *I3\*F* and between them and this first component is greater than 0.9, so they can be considered as essentially the same characteristic. The factor loadings of the numbers of citations (NCit) and publications (NPub) on this first factor are greater than 0.8. NPub, which is the size indicator of output (number of publications), does not load substantially on the second factor which represents impact (number of citations); however, the number of citations (NCit) loads on factor 1 (.802) much more than on factor 2 (.324).

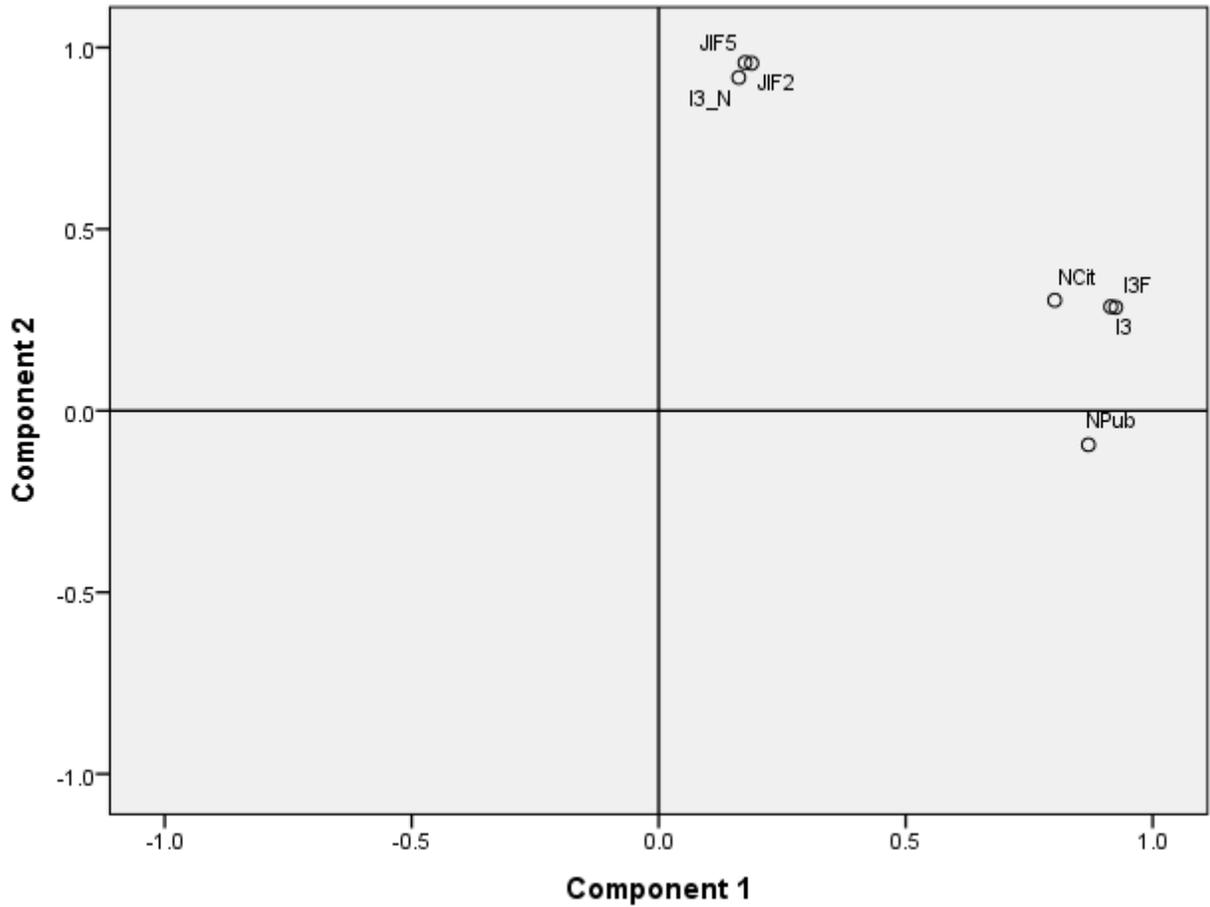

**Figure 2**: Component plot in rotated space of the two main components in the matrix (varimax-rotated PCA) of 10,942 cases (journals) versus seven indicators: total numbers of publications (NPub), citations (NCit), JIF2, JIF5, non-normalized *I3\**-values (*I3\**), field-normalized *I3\**-values (*I3\*F*), and *I3/N\** for the non-normalized case (*I3\*/N*).[3]

---

[3] The asterisk is an illegal character in a variable name or label in SPSS, and therefore not included in the plots.

**Table 3**: Rotated factor matrix of the seven indicators plotted in Figure 1.

**Rotated Component Matrix**[a]

| Indicator | Component 1 | Component 2 |
|---|---|---|
| *I3\*F* | **.925** | .284 |
| *I3\** | **.915** | .286 |
| NPub | **.870** | -.094 |
| NCit | **.802** | .304 |
| JIF5 | .175 | **.958** |
| JIF2 | .188 | **.957** |
| *I3\*/N* | .162 | **.917** |

Notes. Extraction Method: Principal Component Analysis. Rotation Method: Varimax with Kaiser Normalization. The indicators are: total numbers of publications (NPub); citations (NCit); JIF2; JIF5; non-normalized *I3\**-values (*I3\**); field-normalized *I3\**-values (*I3\*F*); and scaled *I3\** for the non-normalized case (*I3\*/N*).

[a.] Rotation converged in 3 iterations.

**Table 4**: Spearman rank-order correlations between the variables listed in Table 3

|   | *NCit* | *JIF2* | *JIF5* | *NPub* | *I3\** | *I3\*F* | *I3\*/N* |
|---|---|---|---|---|---|---|---|
| NCit | 1.000 | .766** | .776** | .719** | .816** | .802** | .706** |
| *JIF2* |  | 1.000 | .924** | .444** | .668** | .638** | .882** |
| *JIF5* |  |  | 1.000 | .417** | .635** | .623** | .848** |
| *NPub* |  |  |  | 1.000 | .920** | .861** | .420** |
| *I3\** |  |  |  |  | 1.000 | .940** | .697** |
| *I3\*F* |  |  |  |  |  | 1.000 | .683** |
| *I3\*/N* |  |  |  |  |  |  | 1.000 |

Notes. The indicators are: total numbers of publications (NPub); citations (NCit); JIF2; JIF5; non-normalized *I3\**-values (*I3\**); field-normalized *I3\**-values (*I3\*F*); and scaled *I3\** for the non-normalized case (*I3\*/N*).

The correlations in Table 4 are all statistically significant ($p<.01$). Note that the number of journals is large ($n = 10,942$) and that significance is therefore less meaningful. However, it can be noted that JIF2 and JIF5 correlate with publication count (NPub) at an observably lower level (0.44 and 0.42) than *I3\** and *I3\*F* (0.92 and 0.86). Obviously, size-normalization (dividing by *n*) does not completely remove the effect of size. This is in accordance with the recently published conclusions of Antonoyiannakis (2018). *I3/N* can also be considered as a mean and thus a parametric statistic.

**Table 5**: 25 journals ranked on non-normalized *I3\** values (*I3\**), field-normalized values (*I3\*F*), and *I3\*/N* (non-normalized). For the full list see http://www.leydesdorff.net/*I3*/ranking.htm .[1]

| JOURNAL | I3* | JOURNAL | I3*F | JOURNAL | I3* / N |
|---|---|---|---|---|---|
| *PLOS One* | 78,733 | *PLOS One* | 53,570.26 | Nat. Rev. Drug Discov. | 90.8 |
| J. Am. Chem. Soc. | 55,786 | Nature | 27,397.23 | Physiol. Rev. | 76.8 |
| Nature | 52,888 | Phys. Rev. Lett. | 24,909.74 | Nat. Rev. Genet. | 72.5 |
| Proc. Natl. Acad. Sci. U. S. A. | 47,041 | Adv. Mater. | 23,741.25 | Prog. Mater. Sci. | 71.9 |
| Nat. Commun. | 46,762 | Nat. Commun. | 21,689.35 | Nat. Rev. Mol. Cell Biol. | 70.6 |
| Science | 41,946 | Science | 21,493.38 | Nat. Rev. Cancer | 69.8 |
| Angew. Chem.-Int. Edit. | 40,572 | Proc. Natl. Acad. Sci. U. S. A. | 21,204.46 | Chem. Rev. | 63.7 |
| Adv. Mater. | 34,435 | J. Am. Chem. Soc. | 20,435.29 | Nat. Rev. Neurosci. | 63.1 |
| Phys. Rev. Lett. | 30,549 | J. Mater. Chem. A | 18,323.90 | Nat. Rev. Immunol. | 62.1 |
| ACS Nano | 29,284 | Nano Lett. | 16,905.86 | N. Engl. J. Med. | 62.0 |
| J. Mater. Chem. A | 28,260 | ACS Nano | 16,608.79 | Nature | 61.4 |
| Chem. Commun. | 26,209 | Phys. Rev. B | 16,176.17 | Living Rev. Relativ. | 57.7 |
| Nano Lett. | 24,717 | Appl. Phys. Lett. | 16,077.51 | Chem. Soc. Rev. | 56.8 |
| ACS Appl. Mater. Interfaces | 24,407 | Angew. Chem.-Int. Edit. | 15,136.02 | Lancet | 55.8 |
| RSC Adv. | 24,326 | Opt. Express | 14,893.89 | Rev. Mod. Phys. | 55.7 |
| Cell | 22,993 | Org. Lett. | 14,819.56 | Cell Stem Cell | 54.6 |
| N. Engl. J. Med. | 21,874 | Energy Environ. Sci. | 14,565.50 | Nat. Photonics | 54.4 |
| Chem. Soc. Rev. | 21,576 | RSC Adv. | 14,479.96 | Nature Genet. | 54.2 |
| Sci Rep | 20,098 | Cell | 13,689.86 | Prog. Polym. Sci. | 53.5 |
| Nanoscale | 19,631 | ACS Appl. Mater. Interfaces | 13,678.24 | Nat. Biotechnol. | 53.2 |
| Astrophys. J. | 19,130 | Anal. Chem. | 12,548.23 | Psychol. Sci. Public Interest | 52.8 |
| Phys. Rev. B | 18,831 | Phys. Chem. Chem. Phys. | 12,459.76 | Cell | 52.7 |
| Chem. Rev. | 17,889 | Nanoscale | 12,418.65 | Nat. Med. | 52.5 |
| Energy Environ. Sci. | 17,151 | Astrophys. J. | 12,269.03 | Nat. Mater. | 51.3 |
| Phys. Rev. D | 16,893 | Chem. Soc. Rev. | 11,710.45 | Cancer Cell | 51.2 |

---

[1] For the sake of readability, journal names in this table follow the ISO Abbreviations instead of those in Journal Citation Reports.

Table 5 shows the ranking of the 25 journals with the greatest index values for each of *I3\**, field-normalized *I3\*F,* and *I3\*/N*, respectively. (The full listing is available at http://www.leydesdorff.net/*I3*/ranking.htm .) The size effect of *PLOS ONE* dominates both the *I3\** and *I3\*F* ranking, but not the third column (*I3\*/N*) which is size-independent because of the division by *N.* Twelve of the 25 titles in this latter column are attributed to journals in the *Nature* publishing group indicating the high quality of this portfolio. Note that *Science*, which occupies sixth position in the first two columns, drops to 29$^{th}$ position on the size-independent indicator. *PLOS One* falls much further, to position 2,064.

There may be a disciplinary interaction with normalization: field-normalization seems to affect the leading chemistry journals more than others. The *Journal of the American Chemical Society* (JACS), for example, holds second place on the (left-side) list of *I3\**, but only ninth place on *I3\*F*. By contrast, leading physics journals seem to list higher on the normalized indicator. Perhaps, these relatively well-cited journals in chemistry have a longer-tailed citation distribution than comparable physics journals: normalization (division by *N*) will have a greater effect with increasing values of *N*. As noted, the two indicators are highly correlated overall, but the possibility of a disciplinary, and therefore research-cultural, factor will need further elucidation.

*3.2. The Social Sciences Citation Index* (SSCI)

The citation environment of journals listed in the *Social Sciences Citation Index* is very different from that of journals in the SCI-E*.* The SCI-E journals in JCR constitute about 28% (3,105 / 10,942) of the total serial titles, but the total citations to SSCI journals constitute less than 10% of all citations to JCR titles (4,506,510/48,340,046 in our time window). The average yearly total cites (*NCit*) of a journal in SSCI is 1,451.3 compared with 4,417.8 for the combined set. Figure 3 shows the relatively small contribution of the SSCI journals to the citation indices in terms of citations.

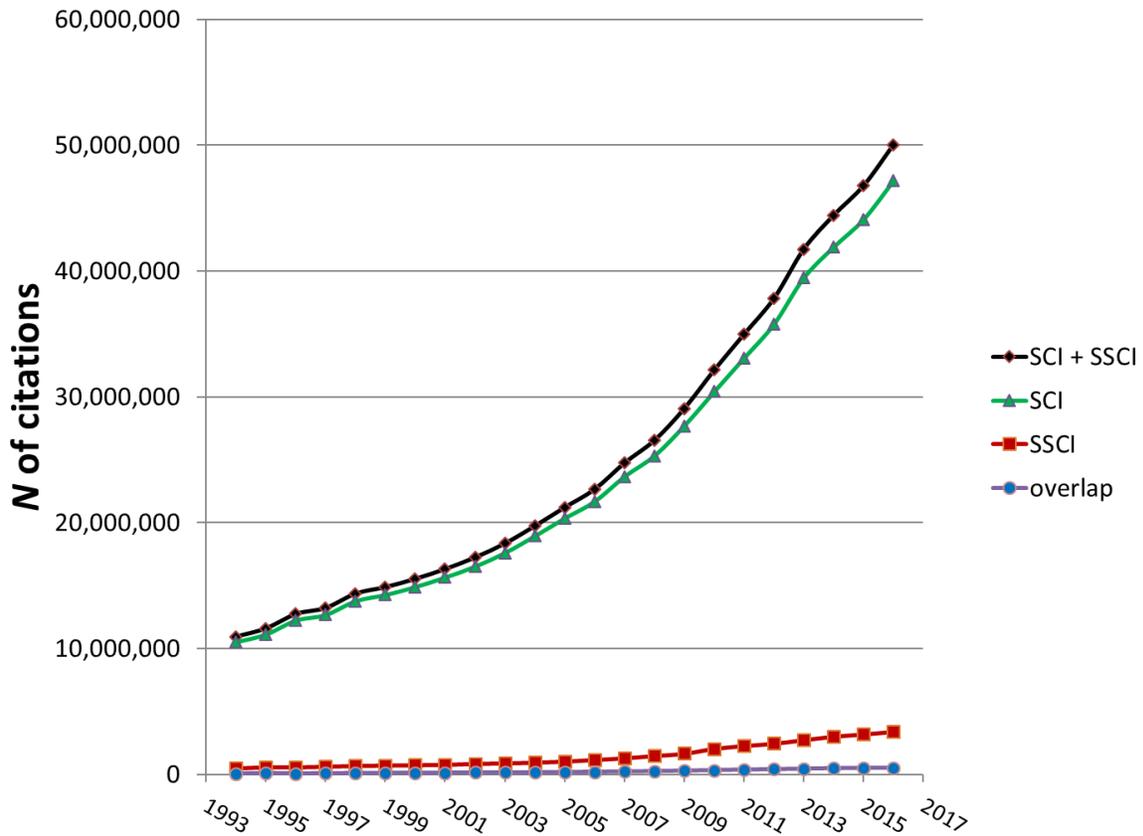

**Figure 3:** Aggregated citation counts for journals in the Web of Science editions for the SCI and SSCI, both separate and combined. Source: Leydesdorff, Wagner, & Bornmann (2018, p. 627).

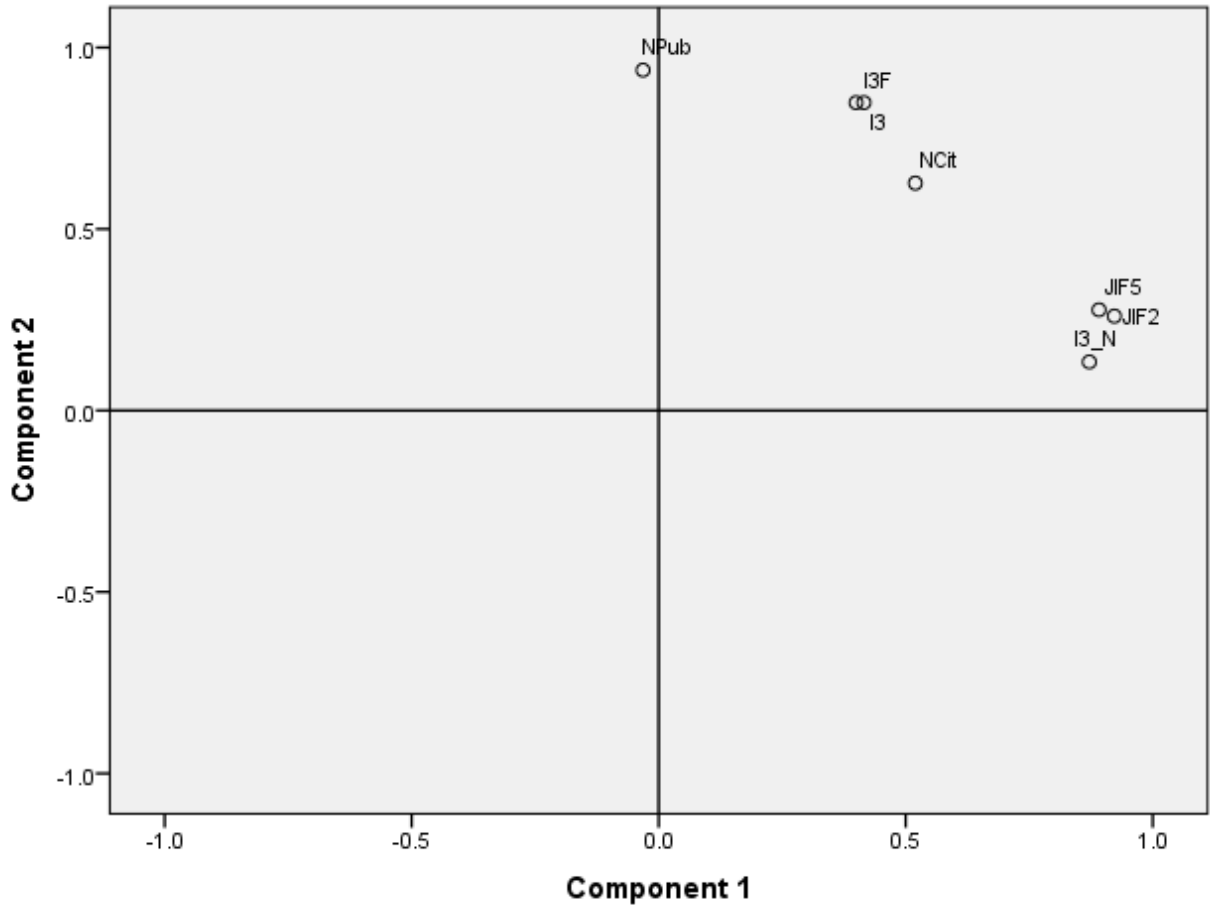

**Figure 4**: Two-component factor plot of seven indicators for 3,105 journals in the SSCI.
Notes: The indicators are: total numbers of publications (NPub); citations (NCit); JIF2; JIF5; non-normalized *I3\**-values (*I3\**); field-normalized *I3*-values (*I3\*F*); and scaled *I3\** for the non-normalized case (*I3\*_N*).

Figure 4 shows the factor plot for the 3,105 journals in SSCI for comparison with the factor plot for the combined sets of SSCI and SCI provided in Figure 1. The main difference is the greater distance between the number of citations (NCit) and the number of publications (NPub). The correlation between both indicators is smaller in SSCI (.633) than in SCI (.719) (Tables 4 and 7, respectively). Consequently, NPub and NCit are distanced in Figure 4 and the order of the two factors is reversed. Nonetheless, these two factors together still explain 84% of the variance.

**Table 6**: Rotated factor matrix of the seven indicators plotted in Figure 4.

| Indicator | Component 1 | Component 2 |
|---|---|---|
| JIF2 | **.922** | .260 |
| JIF5 | **.891** | .277 |
| *I3*/N* | **.872** | .134 |
| NPub | -.032 | **.938** |
| *I3** | .416 | **.848** |
| *I3*F* | .399 | **.848** |
| NCit | **.519** | **.626** |

Notes: Extraction Method: Principal Component Analysis. Rotation Method: Varimax with Kaiser Normalization. The indicators are: total numbers of publications (NPub); citations (NCit); JIF2; JIF5; non-normalized *I3**-values (*I3**); field-normalized *I3**-values (*I3*F*); and scaled *I3** for the non-normalized case (*I3*/N*).

a. Rotation converged in 3 iterations.

**Table 7**: Spearman rank-order correlations among the seven indicators under study for 3,105 journals in the SSCI.

| | *NCit* | *JIF2* | *JIF5* | *NPub* | *I3\** | *I3\*F* | *I3\*/N* |
|---|---|---|---|---|---|---|---|
| *NCit* | 1.000 | .799** | .820** | **.633**** | .775** | .742** | .746** |
| *JIF2* | | 1.000 | .881** | .412** | .644** | .626** | .846** |
| *JIF5* | | | 1.000 | .397** | .616** | .605** | .797** |
| *NPub* | | | | 1.000 | .904** | .786** | .403** |
| *I3\** | | | | | 1.000 | .893** | .704** |
| *I3\*F* | | | | | | 1.000 | .693** |
| *I3\*/N* | | | | | | | 1.000 |

Notes: The indicators are: total numbers of publications (NPub); citations (NCit); JIF2; JIF5; non-normalized *I3**-values (*I3**); field-normalized *I3**-values (*I3*F*); and scaled *I3** for the non-normalized case (*I3*/N*).

If we focus on a specific journal category of SSCI, such as the 83 journals in Information & Library Science, the difference in citation cultures between SCI and SSCI outcomes is further emphasized. Alternatively, if we focus on a narrow specialism in the natural sciences, such as Spectroscopy with 41 journals, we find that the distinction between the two components is even more pronounced than for the full set of 10,942 journals. Table 8 juxtaposes the rotated factor matrices showing these differences numerically.

**Table 8**: Rotated factor matrices for two specialist WoS Categories, one each from SSCI and SCI.

| Library and Information Science, 83 journals | | | Spectroscopy, 41 journals | | |
|---|---|---|---|---|---|
| **Rotated Component Matrix**[a] | | | **Rotated Component Matrix**[a] | | |
| Indicator | Component | | Indicator | Component | |
| | 1 | 2 | | 1 | 2 |
| JIF5 | **0.959** | 0.170 | *I3\** | **0.982** | |
| JIF2 | **0.927** | 0.215 | *I3\*F* | **0.962** | |
| *I3\*/N* | **0.799** | 0.265 | NPub | **0.959** | -0.113 |
| **NCit** | **0.757** | *0.527* | NCit | **0.771** | 0.167 |
| NPub | | **0.968** | JIF2 | | **0.987** |
| *I3\** | 0.41 | **0.903** | JIF5 | | **0.977** |
| *I3\*F* | 0.444 | **0.861** | *I3\*/N* | | **0.930** |

Notes: Extraction Method: Principal Component Analysis. Rotation Method: Varimax with Kaiser Normalization. The indicators are: total numbers of publications (NPub); citations (NCit); JIF2; JIF5; non-normalized *I3\**-values (*I3\**); field-normalized *I3\**-values (*I3\*F*); and scaled *I3\** for the non-normalized case (*I3\*/N*).

[a.] Rotation converged in 3 iterations.

While the number of publications drives the number of citations in the SCI-E, this appears to be less the case in the SSCI. Size is less important for impact in SSCI than in SCIE. *I3\** correlates with size (NPub) more than with citations (NCit) in the social sciences

**3.5 Comparison with 2009**

It is possible that the results obtained for 2014 were specific for that year, because it is relatively recent and the citation counts were not yet stable. We tested this by repeating the analysis for 2009 data, which was chosen because the Web of Science (version 5) was reorganized in 2008/2009.

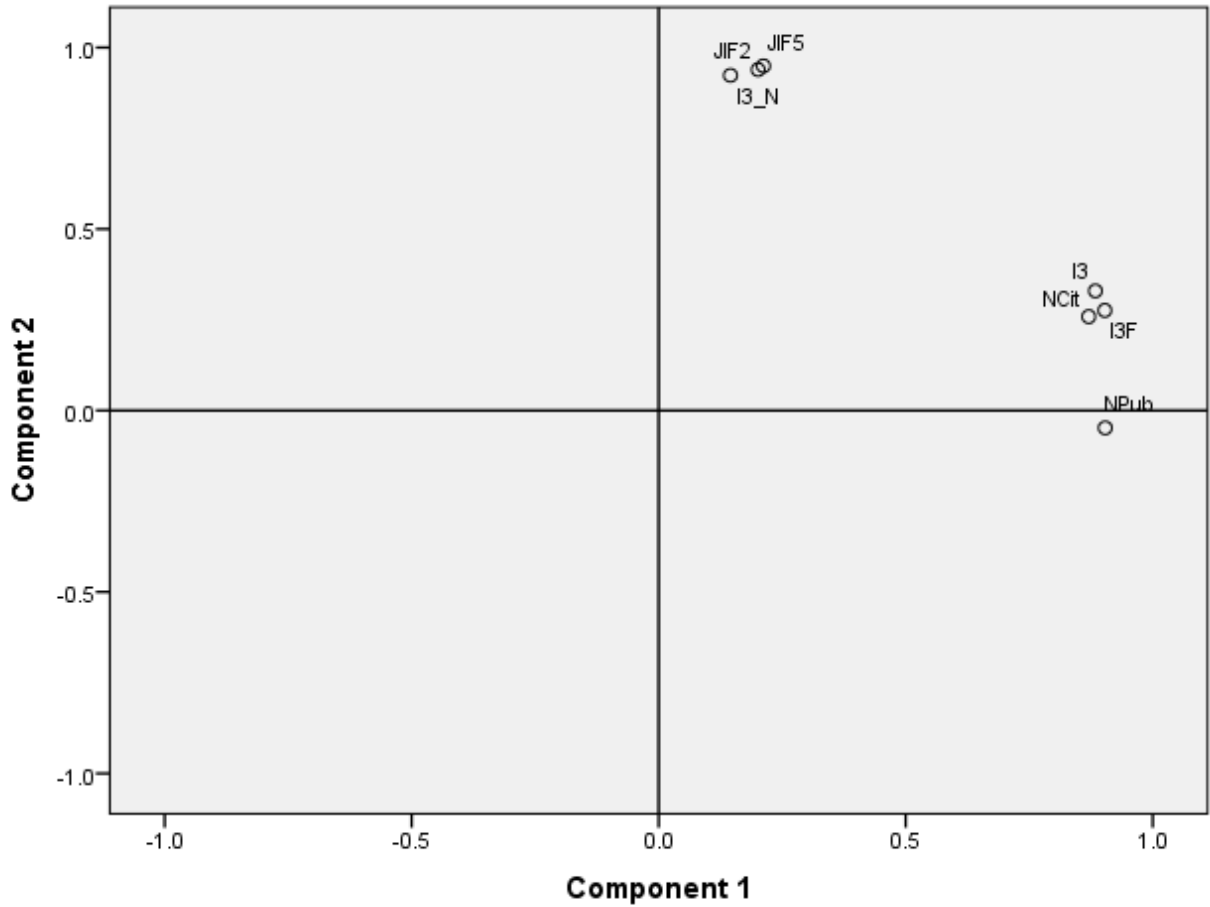

**Figure 5**: Plot of the two first components for 8,994 journals in JCR 2009.

Notes: The indicators are: total numbers of publications (NPub); citations (NCit); JIF2; JIF5; non-normalized *I3\**-values (*I3\**); field-normalized *I3\**-values (*I3\*F*); and scaled *I3\** for the non-normalized case (*I3\*/N*).

Of the 9,216 journals in the combined 2009 sets of JCRs for SCI-E and SSCI, 8,994 journal title abbreviations could automatically be matched between the data from the in-house database and JCR. The two-component plot in Figure 5 shows that the outcome for 2009 data is very similar to that seen with 2014 data (Table 9).

**Table 9**: Rotated factor matrices for full sets in 2009 and 2014.

| JCR 2009: 8,904 journals | | | JCR 2014: 10,942 journals | | |
|---|---|---|---|---|---|
| **Rotated Component Matrix**[a] | | | **Rotated Component Matrix**[a] | | |
| | Component | | | Component | |
| | 1 | 2 | | 1 | 2 |
| **NPub** | **0.904** | -0.048 | *I3*F* | **0.925** | 0.284 |
| *I3*F* | **0.903** | 0.276 | *I3** | **0.915** | 0.286 |
| *I3** | **0.884** | 0.329 | **NPub** | **0.870** | |
| *NCit* | *0.87* | *0.259* | *NCit* | *0.802* | *0.304* |
| JIF5 | 0.212 | **0.949** | JIF5 | 0.175 | **0.958** |
| JIF2 | 0.201 | **0.939** | JIF2 | 0.188 | **0.957** |
| *I3*/N* | 0.145 | **0.923** | *I3*/N* | 0.162 | **0.917** |

Notes: Extraction Method: Principal Component Analysis. Rotation Method: Varimax with Kaiser Normalization. The indicators are: total numbers of publications (NPub); citations (NCit); JIF2; JIF5; non-normalized *I3**-values (*I3**); field-normalized *I3**-values (*I3*F*); and scaled *I3** for the non-normalized case (*I3*/N*).

[a.] Rotation converged in 3 iterations.

Two factors explain 88.1% of the variance in 2009 and 87.5% in 2014. Figure 5 shows the 2-component plot for 2009. The results are virtually identical in these two sample years. Thus, the indicator appears to be robust over time.

## 4. Statistics

*PLOS One* was by far the largest journal in 2014 with 30,042 publications. It was followed in this analysis by *RSC Advances* with 8,345 citable items. In terms of total citations, however, *PLOS One* is in eighth place with 332,716 citations. In the same year, *Nature* accrued 617,363 citations to 862 publications. The simple citations/publication (c/p) ratio for *Nature* is 716.3 and for *PLOS One* is 11.1. By comparison, the values of *I3\*/N* are 61.4 for *Nature* and 2.6 for *PLOS One* and, in seeming contradiction to conventional indicators, the (non-normalized) *I3\** values are 78,733 for *PLOS One* and 52,883 for *Nature*.

What do these figures mean, and are the differences statistically and practically significant? One can test the distribution of papers over the classes against the expected numbers. This can be done for the frequencies in the matrix using chi-square statistics, or by a test between means (in the case of *I3\*/N*) using the *z*-test and/or Cohen's **h** for "practical significance."[12] Table 10 shows various options for testing observed values against expected ones; Table 11 generalizes this to the possibility to test any two distributions against each other. As empirical instances, we again use *PLOS ONE* for comparison of observed with expected values (Table 10), and this same journal versus *RSC Advances* in Table 11.

The results of the chi-square tests are statistically significant ($p < .001$), both when comparing *PLOS ONE* with the expectation, and *PLOS ONE* with *RSC Advances*. One can summarize the results of the chi-square *ex post* using Cramèr's *V* which conveniently ranges from zero to one. In this case, Cramèr's $V = 0.27$ in Table 10 and Cramèr's $V = 0.05$ in Table 11. In other words, the differences between the expected and observed percentile-rank distribution is more than five times larger than the corresponding differences between *PLOS ONE* and *RSC Advances*. (The template provides these values automatically.) The results of the chi-square based on testing the *I3\** values (in columns *g* and *h* in both Tables 9 and 10) are provided in column *k* at the bottom.

---

[12] Cohen's **h** tests proportions against each other for each row using $\mathbf{h} = 2*(\arcsin\sqrt{p_{obs}} - \arcsin\sqrt{p_{exp}})$ (Cohen, 1988, pp. 180 ff.), whereas Cohen's **w** first sums over the rows and then takes the square root (Cohen, 1988, pp. 216f.): $\mathbf{w} = \sqrt{\sum_{i=1}^{m} \frac{(p(observed) - p(expected))^2}{p(expected)}}$.

**Table 10**: Comparison of *PLOS One* with expected values

| PLOS One | observed | expected | classes | observed | expected | I3* | I3*_exp | standardized residuals of the χ2 | | χ2 |
|---|---|---|---|---|---|---|---|---|---|---|
| (a) | (b) | (c) | (d) | (e) | (f) | (g) | (h) | (i) | (j) | (k) |
| top-1% | 91 | 300.42 | **99-100** | 91 | 300.42 | 9100 | 30042 | -6.10 | p<.001 | 7498.42 |
| top10% | 2545 | 3004.2 | **90-98** | 2454 | 2703.78 | 24540 | 27038 | 0.82 | n.s. | 135.94 |
| top-50% | 20141 | 15021 | **50-89** | 17596 | 12016.8 | 35192 | 24034 | 4.96 | p<.001 | 4958.68 |
| bottom-50% | 7265 | 15021 | **0-49** | 9901 | 15021 | 9901 | 15021 | -1.18 | n.s. | 282.45 |
| **Sum** | **30042** | **33346.62** | | **30042** | **30042** | **78733** | **75402** | | **χ2 =** | **12,875.49** |
| | | | | | | | | | **df = 3** | **p<.001** |
| | | | | | | | | **Cramèr's V = 0.271** | | |

| PLOS One | I3* / N obs. | I3* /N exp. | p(obs) | p(exp) | z-test | | Cohen's w | Cohen's h |
|---|---|---|---|---|---|---|---|---|
| (a) | (l) | (m) | (o) | (p) | (q) | (r) | (s) | (t) |
| top-1% | **0.303** | **1.000** | 0.0030 | 0.01 | -0.086 | n.s. | 0.005 | -0.090 |
| top10% | **0.817** | **0.900** | 0.0817 | 0.09 | -0.028 | n.s. | 0.001 | -0.030 |
| top-50% | **1.171** | **0.800** | 0.5857 | 0.4 | 0.265 | n.s. | 0.086 | 0.374 |
| bottom-50% | **0.330** | **0.507** | 0.3296 | 0.5 | -0.265 | n.s. | 0.058 | -0.348 |
| **Sum** | **2.62** | **3.20** | 1.0000 | 1 | | | **0.387** | |

| * critical values (higher values than the critical values are statistically significant) | | χ2; df = 3 | z |
|---|---|---|---|
| | **p < 0.001** | 16.266 | 3.291 |
| | **p < 0.01** | 11.345 | 2.576 |
| | **p < 0.05** | 7.815 | 1.96 |

**Table 11**: Comparison of *PLOS One* with *RSC Advances*

| PLOS One vs. RSC Advances (a) | unit 1 (b) | unit2 (c) | Classes (d) | n1 (e) | n2 (f) | I3*_1 (g) | I3*_2 (h) | standardized residuals of the χ2 (i) | (j) | χ2 (k) |
|---|---|---|---|---|---|---|---|---|---|---|
| top-1% | 91 | 30 | **99-100** | 91 | 30 | 9100 | 3000 | 1.196 | n.s. | 9.493 |
| top10% | 2545 | 909 | **90-98** | 2454 | 879 | 24540 | 8790 | 4.621 | p <.001 | 141.686 |
| top-50% | 20141 | 5919 | **50-89** | 17596 | 5010 | 35192 | 10020 | -2.802 | p < .01 | 52.113 |
| bottom-50% | 7265 | 1577 | **0-49** | 9901 | 2516 | 9901 | 2516 | -3.404 | p <.001 | 76.881 |
| Sum | 30042 | 8435 | | 30042 | 8435 | 78733 | 24326 | | **χ2 =** | **280.173** |
| | | | | | | | | | **df = 3** | **p<.001** |
| | | | | | | | | | | **Cramèr's V = 0.0521** |

| PLOS One vs. RSC Advances (a) | I3*/N unit1 (l) | I3*/N unit2 (m) | p1 (o) | p2 (p) | z-test (q) | (r) | Cohen's w (s) | Cohen's h |
|---|---|---|---|---|---|---|---|---|
| top-1% | 0.303 | 0.356 | 0.003 | 0.004 | 0.765 | n.s. | 0.000 | 0.009 |
| top10% | 0.817 | 1.042 | 0.082 | 0.104 | 6.498 | P <.001 | 0.005 | 0.078 |
| top-50% | 1.171 | 1.188 | 0.586 | 0.594 | 1.358 | n.s. | 0.000 | 0.017 |
| bottom-50% | 0.330 | 0.298 | 0.330 | 0.298 | -5.432 | P <.001 | 0.003 | -0.067 |
| Sum | 2.621 | 2.884 | **1.000** | **1.000** | | | **0.091** | |

| * critical values | | χ2; df = 3 |
|---|---|---|
| | p < 0.001 | 16.266 |
| | p < 0.01 | 11.345 |
| | p < 0.05 | 7.815 |

z

While the chi-square statistic provides a test for comparing the entire distributions (two vectors of four classes), the decomposition of chi-square into standardized residuals [$\frac{(observed-expected)}{\sqrt{expected}}$] provides us with a statistic for each class. Standardized residuals can be considered as *z*-values: they are significant at the 5% level if the absolute value is larger than 1.96, 1% for an absolute value > 2.576, and 1‰ for an absolute value > 3.291 (Sheskin, 2011, at p. 672).

Furthermore, the residuals are signed and indicate (in Table 10, for example) that *PLOS One* scores are significantly below expectation in the top-1% class (*p*<.001), but above expectation in the top-50% class (*p*<.001). The overall distribution over the percentile classes (including the vertical direction of columns *e* and *f*) is statistically significant at the 1‰ level: the journal as a whole performs significantly below expectation in terms of *I3\**. (Note that each of the four decompositions in column *l* is based on two observations, since eight cells are used in the computation of the chi-square.)

In Table 11, *RSC Advances* scores statistically significantly higher than *PLOS One* in the top-10% (column *l*), but not statistically significant below *PLOS One* in the lower-ranked classes. Tables 10b and 11b add the statistics for *I3\*/N*. The division by *N* makes all the frequencies relative. Since these relative frequencies can also be considered as proportions, one can *z*-test for difference in proportions (Sheskin, 2011, pp. 656f.) or also compute an effect size using Cohen's **w** (1988, at p. 216; Leydesdorff, Bornmann, & Mingers, in press).

The *z*-values in column *q* of Table 10 show (in column *r*) that *PLOS One* scores above expectation in the percentile class between 50 and 89, but this value is not statistically significant. *PLOS One* scores non-significantly below expectation in the top-1% and even more so in the top-10% and bottom-50%.

These results may come as no surprise, but in cases other than *PLOS One* may offer less intuitive results about the status of a journal. For example, specification of the differences between *RSC Advances* and *Nature* in terms of these four classes would be far from obvious. The template available at https://www.leydesdorff.net/*I3*/template.xlsx automatically fills out the numbers and significance levels when the user provides the field-normalized and non-normalized values for top-1%, top-10%, top-50%, and total number of papers in the respective cells.

In order to have information about the significance of the results on the basis of effect sizes (Cohen, 1988; Schneider, 2013; Wasserstein & Lazar, 2016; Williams & Bornmann, 2014), we added Cohen's **h** and **w** for the comparison among proportions as column *s* to Tables 10 and 11. The **w** index is 0.4 in Table 10, and thus the difference between *PLOS ONE* and its expected citation rates in these four categories is meaningful and significant for practical purposes. This is not the case for the difference between the two journals: **w** = 0.1. The values of **h** accord with those of the *z*-test for each of the classes.

It should be kept in mind that these tests on proportions address the size-independent indicator *I3\*/N*. This measure can be used as the expected value of citations of a publication published in the relevant journal. In other words, a paper that is accepted for publication in *RSC Advances* has a significantly greater likelihood of being cited in the overall top-10% than a paper in *PLOS One*. It is also less likely to be cited below the 50% threshold.

## 5. Effects of different weighting schemes

Weighting schemes have a significant effect on the outcome and interpretation of the analysis of categorized data; weighting introduces a level of subjectivity. Using the general scheme of *I3*, *I3* variants can be adapted to the context of the evaluation situation. For example, if the focus is solely on research excellence, the percentile classes reflecting high impact can be provided with a higher weight. Reducing the weighting for higher impact classes would mean that productivity is relatively more emphasized.

What happens if, instead of the logarithmic set, we use the linear set of Mutz & Bornmann (2011) specified in Table 1 or the respective quantile values as used by Leydesdorff & Bornmann (2011)? Our data collection is categorized in four classes, so we can do this with a weight of 6 for the top-1% papers, 4 for the top-10%, 2 for the top-50%, and 1 for the bottom-50%. Mutz & Bornmann (2011) used two additional classes: 5 for the top-5% papers, and the class between 50 and 89 was divided into 75-89 weighted with 3 and the class 50-74 weighted with 2. The analysis is now less sensitive: using a linear scale, the information benefit of *I3\** is considerably reduced.

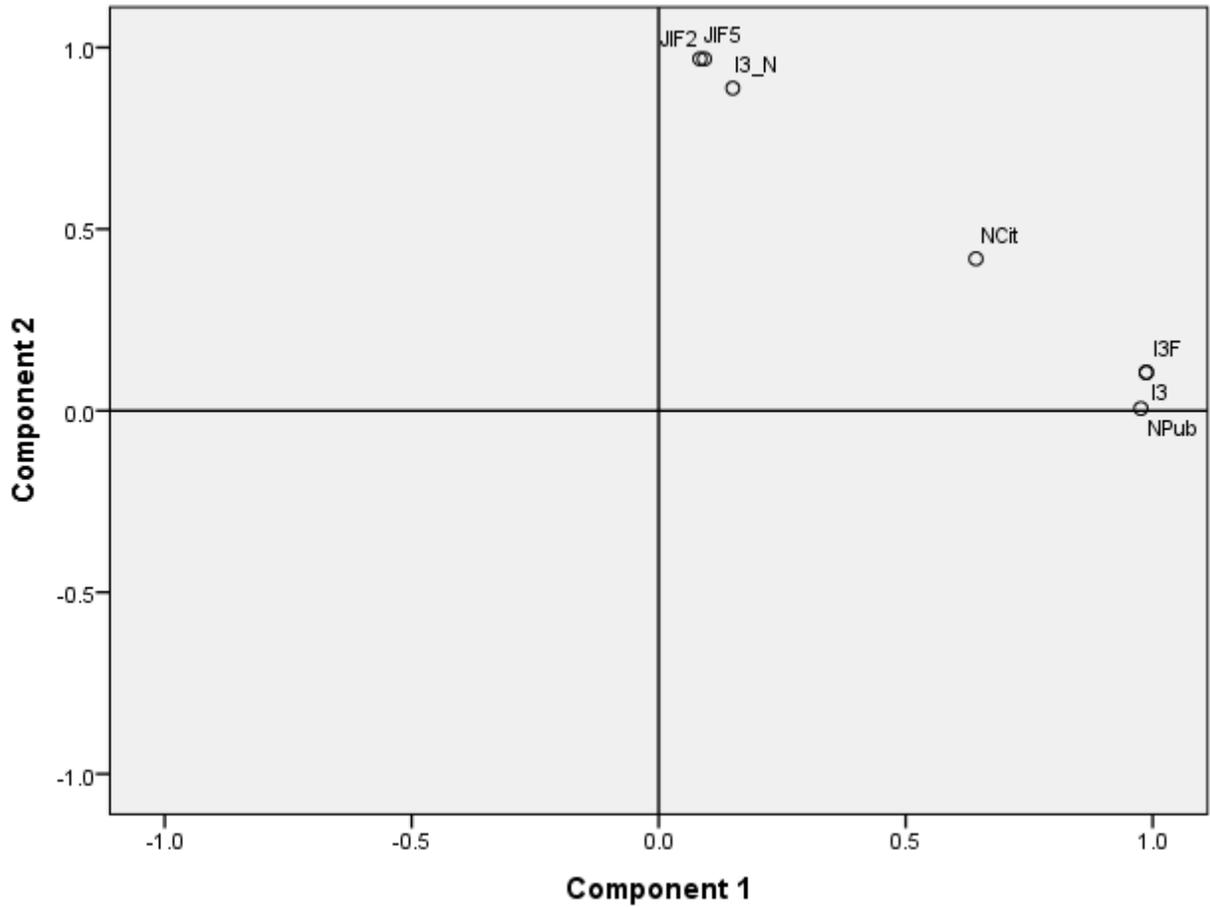

**Figure 6**: Plot of the two main components in the matrix (varimax-rotated PCA) of 10,942 cases (journals). The indicators are: total numbers of publications (NPub); citations (NCit); JIF2; JIF5; non-normalized *I3\**-values (*I3\**); field-normalized *I3\**-values (*I3\*F*); and scaled *I3\** for the non-normalized case (*I3\*/N*).

**Table 12**: Rotated factor matrices of the seven indicators based on replacing *I3\** values with six percentile ranks (Mutz & Bornmann, 2011) and quantile values (Leydesdorff & Bornmann, 2011), respectively.

| **Rotated Component Matrix**[a] | | | | **Rotated Component Matrix**[a] | | |
|---|---|---|---|---|---|---|
| | Component | | | | Component | |
| | 1 | 2 | | | 1 | 2 |
| I3*F | .987 | .106 | | I3* | .981 | .121 |
| I3* | .987 | .105 | | I3*F | .978 | .147 |
| NPub | .976 | .006 | | NPub | .964 | -.001 |
| **NCit** | **.642** | **.418** | | **NCit** | **.660** | **.404** |
| JIF5 | .084 | .969 | | JIF5 | .087 | .963 |
| JIF2 | .093 | .968 | | JIF2 | .096 | .963 |
| I3*/N | .150 | .888 | | I3* /N | .184 | .820 |
| Extraction Method: Principal Component Analysis. | | | | Extraction Method: Principal Component Analysis. | | |
| Rotation Method: Varimax with Kaiser Normalization.[a] | | | | Rotation Method: Varimax with Kaiser Normalization. | | |
| a. Rotation converged in 3 iterations. | | | | a. Rotation converged in 3 iterations. | | |
| Rotated Factor Matrix using six percentile ranks (Mutz & Bronmann, 2011) | | | | Rotated Factor Matrix using quantile values (Leydesdorff & Bronmann, 2011) | | |

With linear weighting, Figure 6 shows us that *I3\** no longer captures the number of citations, but becomes a size indicator (correlated to <u>*NPub*</u> more than *NCit*). The top-1% papers, for example, are now given a relative value of six instead of one hundred and thus highly skewed, citation frequencies no longer play a strongly differential role in the assessment across higher and lower-ranked percentile classes.

Table 12 shows the comparison between using the six percentile ranks used by Mutz & Bornmann (2011) with using the quantile values 99, 89, 50, and 1 for the four classes (Leydesdorff & Bornmann, 2011). The position of *NCit* is similarly changed in both cases; the coordinate values of *NCit*—boldfaced in Table 12—are slightly lower. Correspondingly, the Pearson correlation of *NCit* and *I3\** declines further from .624 to .613 ($p<.01$).

Recall a similar effect for the social sciences (SSCI compared with SCI) particularly when we focused on the 83 journals in the LIS category. The reason in that case was because of a difference in the data, but in the general case the reason is the (mis)specification of a model which does not give appropriate attention to the skew in the distribution.

## 6. Summary and conclusions

We argue in this paper that an indicator can be developed that reflects both impact and output, and that combines the two dimensions of publications and citations into a single measure by using non-parametric statistics. The generic Integrated Impact Indicator *I3* is a sum of weighted publication numbers in different percentile classes. The indicator can be used very flexibly with a range of percentile classes and weights. Depending on the chosen parameters, *I3* can be made more output- or more impact-orientated. In this study, we introduced *I3\** = *I3*(99-100, 90-10, 50-2, 0-1) which categorises and weights papers published in the higher citation impact range in a more informed way, given the distribution skew, than the indicator proposed by Mutz & Bornmann (2011) and the quantile-based approach elaborated by Leydesdorff & Bornmann (2011).

*I3\** can be size-normalized by dividing the value by the original number of publications, to obtain a secondary indicator that expresses the expected contribution made by a single paper given the journal's characteristics. The size-normalized and size-independent indicators can be considered as relating to two nearly orthogonal axes. When we consider the relationship between conventional journal indicators and these new indicators, we see that *I3\** correlates strongly with both the total number of citations and publications, whereas *I3\*/N* correlates with size-independent indicators such as JIF.

The Journal Impact Factor developed by Garfield and Sher (1963) was originally intended as a journal statistic of value to publishers and librarians for portfolio management. It was not intended for research evaluation, but it has in fact been increasingly employed for this purpose and mistakenly used as a benchmark for individual researchers and their research output. An average citation rate of two (JIF2) or five (JIF5) years is not representative of the journal as a whole. The JIF can be used as one indicator of the reputation or status of a journal, subject to appropriate contextual considerations, but it cannot be used as an impact value for single papers (Pendlebury & Adams, 2012; Bornmann & Williams, 2017; Leydesdorff, Wouters, & Bornmann, 2016; cf. Waltman & Traag, 2018).

Can the *I3\** indicator be compared with the *h*-index? Only to the extent that the measurement of output and impact are combined into a single number in both indicators. However, the *h*-index is mathematically inconsistent; it overrides disciplinary-specific cultural and other considerations, and observed values cannot be tested systematically against expected ones. By contrast, *I3\** can be analyzed using various statistical tests or power analysis depending on the context in which one wishes to use the indicator. Furthermore, *I3\** does not provide only one single value like the *h*-index, but gives an additional four reference values with performance information in different impact classes. This information can be compared with expected values and between different publication sets (e.g., of two or more institutions). Thus, *I3\** can be used as a single number (e.g., for policy purposes), but it can also be decomposed into the contributions of the percentile rank classes (e.g., the top-10% group). Importantly, one is able to specify error terms on the basis of statistics.

The versatility of *I3\** is illustrated in a spreadsheet in Excel containing a template for the computation at https://www.leydesdorff.net/i3/template.xlsx. The $P_{\text{top 10\%}}$ and $PP_{\text{top 10\%}}$ indicators have become established as quasi-standard indicators in professional bibliometrics, especially when research institutions are compared (Waltman *et al*., 2012). The use of these percentile-based indicators is recommended, for instance in the Leiden Manifesto, which included ten guiding principles for research evaluation (Hicks *et al*., 2015).[13] It is an advantage of the *I3\** indicator—which is a percentile-based indicator—that it integrates the top-1% with the top-10% information and combines them with information about other percentile classes. Thus, one provides a broader picture by using *I3\** as indicator compared to $P_{\text{top 10\%}}$ and $PP_{\text{top 10\%}}$.

The almost weekly invention of a new *h*-type indicator signals that many innovative analysts are not aware of a central problem with bibliometric data, shared with other forms of collected data, that indicators necessarily generate error both in source measurement and through analytical methodology (Leydesdorff, Wouters, & Bornmann, 2016, pp. 2144f.). Consequently, one should not underestimate the need to elaborate, test, and report on algorithms and their analytics, both empirically *and* statistically. Elegance on purely mathematical (that is, *a priori*) grounds is not a sufficient claim for scientometric utility (Ye & Leydesdorff, 2014).

---

[13] As explained above, *I3*(90-1) is the notation for $P_{\text{top 10\%}}$ whereas $PP_{\text{top 10\%}}$ can be written as *I3*(90-1)/*N*.

# 7. Perspectives for further research

The convergent validity of different (field-normalized) indicators can be investigated by comparing the indicators with assessments by peers (Bornmann *et al*., 2019). Peer assessments of papers published in the biomedical area are available in the F1000Prime database (see https://f1000.com/prime). High correlations between quantitative and qualitative assessments signal the convergent validity of bibliometric indicators which should be preferred in the practice of research evaluation. Bornmann & Leydesdorff (2013) have correlated different indicators with assessments by peers provided in the F1000Prime database. The results showed, for instance, that "Percentile in Subject Area achieves the highest correlation with F1000 ratings" (p. 286). In a follow-up study, *I3\** indicators are investigated with a similar design to investigate whether these new indicators also have convergent validity (Bornmann *et al*., in press).

**Acknowledgements**
The bibliometric data used in this paper are from an in-house database developed and maintained in collaboration with the Max Planck Digital Library (MPDL, Munich) of the Max Planck Society, and derived from the Science Citation Index Expanded (SCI-E), the Social Sciences Citation Index (SSCI), and the Arts and Humanities Citation Index (AHCI) prepared by Clarivate

Analytics (Philadelphia, Pennsylvania, USA). We are also grateful to ISI/Clarivate Analytics for providing one of us with JCR data.

**References**

Ahlgren, P., Persson, O., & Rousseau, R. (2014). An approach for efficient online identification of the top-k percent most cited documents in large sets of Web of Science documents. *ISSI Newsletter, 10*(4), 81-89.

Alberts, B. (2013). Impact factor distortions. *Science, 340*(6134), 787-787.

Antonoyiannakis, M. (2018). Impact Factors and the Central Limit Theorem: Why citation averages are scale dependent. *Journal of Informetrics, 12*(4), 1072-1088.

Archambault, É., & Larivière, V. (2009). History of the journal impact factor: Contingencies and consequences. *Scientometrics, 79*(3), 635-649.

Baumgartner, S., & Leydesdorff, L. (2014). Group-Based Trajectory Modeling (GBTM) of Citations in Scholarly Literature: Dynamic Qualities of "Transient" and "Sticky Knowledge Claims". *Journal of the American Society for Information Science and Technology, 65*(4), 797-811.

Bensman, S. J. (2007). Garfield and the impact factor. *Annual Review of Information Science and Technology, 41*(1), 93-155.

Bornmann, L. (2014). How are excellent (highly cited) papers defined in bibliometrics? A quantitative analysis of the literature. *Research Evaluation, 23*(2), 166–173.

Bornmann, L., De Moya Anegón, F., & Leydesdorff, L. (2012). The new Excellence Indicator in the World Report of the SCImago Institutions Rankings 2011. *Journal of Informetrics, 6*(2), 333-335. doi: 10.1016/j.joi.2011.11.006

Bornmann, L., & Leydesdorff, L. (2013). The validation of (advanced) bibliometric indicators through peer assessments: A comparative study using data from InCites and F1000. *Journal of Informetrics, 7*(2), 286-291. doi: 10.1016/j.joi.2012.12.003.

Bornmann, L., & Mutz, R. (2011). Further steps towards an ideal method of measuring citation performance: The avoidance of citation (ratio) averages in field-normalization. *Journal of Informetrics, 5*(1), 228-230.

Bornmann, L., Mutz, R., & Daniel, H.-D. (2008). Are there better indices for evaluation purposes than the *h* index? A comparison of nine different variants of the *h* index using data from biomedicine. *Journal of the American Society for Information Science and Technology, 59*(5), 830-837. doi: 10.1002/asi.20806.

Bornmann, L., Mutz, R., Marx, W., Schier, H., & Daniel, H.-D. (2011). A multilevel modelling approach to investigating the predictive validity of editorial decisions: do the editors of a high profile journal select manuscripts that are highly cited after publication? *Journal of the Royal Statistical Society: Series A (Statistics in Society), 174*(4), 857-879.

Bornmann, L., Mutz, R., Hug, S. E., & Daniel, H.-D. (2011). A multilevel meta-analysis of studies reporting correlations between the h index and 37 different h index variants. *Journal of Informetrics, 5*(3), 346-359.

Bornmann, L., Tekles, A., & Leydesdorff, L. (in press). The convergent validity of several (field-normalized) bibliometric indicators: How well does I3 perform for impact measurement? *Scientometrics*.

Bornmann, L., & Williams, R. (2017). Can the journal impact factor be used as a criterion for the selection of junior researchers? A large-scale empirical study based on ResearcherID data. *Journal of Informetrics, 11*(3), 788-799. doi: 10.1016/j.joi.2017.06.001.

Cohen, J. (1988). *Statistical Power Analysis for the Behavioral Sciences (2nd ed.)*. Hillsdale, NJ: Lawrence Erlbaum.

Cozzens, S. E. (1985). Comparing the sciences: Citation context analysis of papers from neuropharmacology and the sociology of science. *Social Studies of Science, 15*(1), 127-153.

Egghe, L. (2008). Mathematical theory of the h-and g-index in case of fractional counting of authorship. *Journal of the American Society for Information Science and Technology, 59*(10), 1608-1616.

Egghe, L., & Rousseau, R. (1990). *Introduction to Informetrics*. Amsterdam: Elsevier.

Frandsen, T. F., & Rousseau, R. (2005). Article impact calculated over arbitrary periods. *Journal of the American Society for Information Science and Technology, 56*(1), 58-62.

Garfield, E. (1955). Citation Indexes for Science: A New Dimension in Documentation through Association of Ideas. *Science, 122*(3159), 108-111.

Garfield, E. (1972). Citation Analysis as a Tool in Journal Evaluation. *Science 178*(Number 4060), 471-479.

Garfield, E. (1975). The "obliteration phenomenon" in science—and the advantage of being obliterated. *Current Contents, December 22, #51/52*, 396–398.

Garfield, E. (1979). Is citation analysis a legitimate evaluation tool? *Scientometrics, 1*(4), 359-375.

Garfield, E. (1987). 100 citation classics from the Journal of the American Medical Association. *Jama, 257*(1), 52-59.

Garfield, E. (2003). The Meaning of the Impact Factor. *Revista Internacional de Psicologia Clinica y de la Salud, 3*(2), 363-369.

Garfield, E. (2006). The history and meaning of the journal impact factor. *JAMA, 295*(1), 90-93.

Garfield, E., & Sher, I. H. (1963). New factors in the evaluation of scientific literature through citation indexing. *American Documentation, 14*(3), 195-201.

Gross, P. L. K., & Gross, E. M. (1927). College libraries and chemical education. *Science, 66*(No. 1713 (Oct. 28, 1927)), 385-389.

Hazen, A. (1914). Storage to be Provided in Impounding Municipal Water Supply. *Transactions of the American Society of Civil Engineers, 77*(1), 1539-1640, at p. 1550.

Hicks, D., Wouters, P., Waltman, L., de Rijcke, S., & Rafols, I. (2015). Bibliometrics: The Leiden Manifesto for research metrics. *Nature, 520*(7548), 429-431.

Hirsch, J. E. (2005). An index to quantify an individual's scientific research output. *Proceedings of the National Academy of Sciences of the USA, 102*(46), 16569-16572.

Jacsó, P. (2009). Five-year impact factor data in the Journal Citation Reports. *Online Information Review, 33*(3), 603-614.

Kreft, G. G., & de Leeuw, E. (1988). The see-saw effect: A multilevel problem? *Quality and Quantity, 22*(2), 127-137.

Leydesdorff, L., & Amsterdamska, O. (1990). Dimensions of Citation Analysis. *Science, Technology & Human Values, 15*(3), 305-335.

Leydesdorff, L., & Bornmann, L. (2011). Integrated Impact Indicators Compared With Impact Factors: An Alternative Research Design With Policy Implications. [Article]. *Journal of the American Society for Information Science and Technology, 62*(11), 2133-2146. doi: 10.1002/asi.21609

Leydesdorff, L., & Bornmann, L. (2012). Percentile Ranks and the Integrated Impact Indicator (*I3*). *Journal of the American Society for Information Science and Technology, 63*(9), 1901-1902.

Leydesdorff, L., & Bornmann, L. (2012). Testing Differences Statistically with the Leiden Ranking. *Scientometrics, 92*(3), 781-783.

Leydesdorff, L., & Milojevic, S. (2015). The Citation Impact of German Sociology Journals: Some Problems with the Use of Scientometric Indicators in Journal and Research Evaluations. [Article]. *Soziale Welt-Zeitschrift Fur Sozialwissenschaftliche Forschung Und Praxis, 66*(2), 193-204. doi: 10.5771/0038-6073-2015-2-193

Leydesdorff, L., Bornmann, L., & Mingers, J. (in press). Statistical Significance and Effect Sizes of Differences among Research Universities at the Level of Nations and Worldwide based on the Leiden Rankings. *Journal of the Association for Information Science and Technology*.

Leydesdorff, L., Bornmann, L., Comins, J., & Milojević, S. (2016). Citations: Indicators of Quality? The Impact Fallacy. *Frontiers in Research Metrics and Analytics, 1*(Article 1). doi: 10.3389/frma.2016.00001

Leydesdorff, L., Bornmann, L., Mutz, R., & Opthof, T. (2011). Turning the Tables on Citation Analysis One More Time: Principles for Comparing Sets of Documents. [Article]. *Journal of the American Society for Information Science and Technology, 62*(7), 1370-1381. doi: 10.1002/asi.21534

Leydesdorff, L., Wagner, C., & Bornmann, L. (2018). Discontinuities in Citation Relations among Journals: Self-organized Criticality as a Model of Scientific Revolutions and Change. *Scientometrics, 116*(1), 623-644. doi: 10.1007/s11192-018-2734-6

Leydesdorff, L., Wouters, P., & Bornmann, L. (2016). Professional and citizen bibliometrics: complementarities and ambivalences in the development and use of indicators-a state-of-the-art report. *Scientometrics, 109*(3), 2129-2150. doi: 10.1007/s11192-016-2150-8

Lundberg, J. (2007). Lifting the crown—citation z-score. *Journal of informetrics, 1*(2), 145-154.

Marchant, T. (2009). An axiomatic characterization of the ranking based on the h-index and some other bibliometric rankings of authors. *Scientometrics, 80*(2), 325-342.

Martyn, J., & Gilchrist, A. (1968). *An Evaluation of British Scientific Journals*. London: Aslib.

McAllister, P. R., Narin, F., & Corrigan, J. G. (1983). Programmatic Evaluation and Comparison Based on Standardized Citation Scores. *IEEE Transactions on Engineering Management, 30*(4), 205-211.

Moed, H. F., & Van Leeuwen, T. N. (1996). Impact factors can mislead. *Nature, 381*(6579), 186.

Mutz, R., & Daniel, H.-D. (2012). Skewed Citation Distributions and Bias Factors: Solutions to two core problems with the journal impact factor. *Journal of Informetrics, 6*(2), 169-176.

Narin, F. (1976). *Evaluative Bibliometrics: The Use of Publication and Citation Analysis in the Evaluation of Scientific Activity*. Washington, DC: National Science Foundation.

Narin, F. (1987). Bibliometric techniques in the evaluation of research programs. *Science and Public Policy, 14*(2), 99-106.

Opthof, T., & Leydesdorff, L. (2010). Caveats for the journal and field normalizations in the CWTS ("Leiden") evaluations of research performance. [Article]. *Journal of Informetrics, 4*(3), 423-430. doi: 10.1016/j.joi.2010.02.003

Pendlebury, D. A. & Adams, J. (2012). Comments on a critique of the Thomson Reuters journal impact factor. *Scientometrics*, *92*, 395-401. doi: 10.1007/s11192-012-0689-6

Ponomarev, I. V., Williams, D. E., Hackett, C. J., Schnell, J. D., & Haak, L. L. (2014). Predicting highly cited papers: A method for early detection of candidate breakthroughs. *Technological Forecasting and Social Change, 81*, 49-55.

Price, D. d. S. (1970). Citation Measures of Hard Science, Soft Science, Technology, and Nonscience. In C. E. Nelson & D. K. Pollock (Eds.), *Communication among Scientists and Engineers* (pp. 3-22). Lexington, MA: Heath.

Robinson, W. D. (1950). Ecological correlations and the behavior of individuals. *American Sociological Review, 15*, 351-357.

Schiffman, S. S., Reynolds, M. L., & Young, F. W. (1981). *Introduction to multidimensional scaling: theory, methods, and applications*. New York / London: Academic Press.

Schneider, J. W. (2013). Caveats for using statistical significance tests in research assessments. *Journal of Informetrics, 7*(1), 50-62.

Seglen, P. O. (1992). The Skewness of Science. *Journal of the American Society for Information Science, 43*(9), 628-638.

Seglen, P. O. (1997). Why the impact factor of journals should not be used for evaluating research. *British Medical Journal, 314*, 498-502.

Sher, I. H., & Garfield, E. (1965). *New tools for improving and evaluating the effectiveness of research.* Paper presented at the Second conference on Research Program Effectiveness, July 27-29, Washington, DC.

Sheskin, D. J. (2011). *Handbook of Parametric and Nonparametric Statistical Procedures (5th Edition)*. Boca Raton, FL: Chapman & Hall/CRC.

Tahamtan, I., & Bornmann, L. (2018). Creativity in Science and the Link to Cited References: Is the Creative Potential of Papers Reflected in their Cited References? *Journal of Informetrics, 12*(3), 906-930.

Tijssen, R. J. W., Visser, M. S., & Van Leeuwen, T. N. (2002). Benchmarking international scientific excellence: are highly cited research papers an appropriate frame of reference? *Scientometrics, 54*(3), 381-397.

Van Raan, A. F. J. (2004). Sleeping beauties in science. *Scientometrics, 59*(3), 467-472.

Waltman, L., & Schreiber, M. (2013). On the calculation of percentile-based bibliometric indicators. *Journal of the American Society for Information Science and Technology, 64*(2), 372-379.

Waltman, L., & Traag, V. A. (2017). Use of the journal impact factor for assessing individual articles need not be wrong. *arXiv preprint arXiv:1703.02334*.

Waltman, L., & Van Eck, N. J. (2012). The inconsistency of the h-index. *Journal of the American Society for Information Science and Technology, 63*(2), 406-415.

Waltman, L., Calero-Medina, C., Kosten, J., Noyons, E., Tijssen, R. J., Eck, N. J., . . . Wouters, P. (2012). The Leiden Ranking 2011/2012: Data collection, indicators, and interpretation. *Journal of the American Society for Information Science and Technology, 63*(12), 2419-2432.

Wasserstein, R. L., & Lazar, N. A. (2016). The ASA's statement on p-values: context, process, and purpose. *The American Statistician, 70*(2), 129-133.

Williams, R., & Bornmann, L. (2014). The substantive and practical significance of citation impact differences between institutions: Guidelines for the analysis of percentiles using effect sizes and confidence intervals. In Y. Ding, R. Rousseau & D. Wolfram (Eds.), *Measuring Scholarly Impact: Methods and practice* (pp. 259-281). Heidelberg: Springer.

Ye, F. Y., & Leydesdorff, L. (2014). The "Academic Trace" of the Performance Matrix: A Mathematical Synthesis of the h-Index and the Integrated Impact Indicator (*I3*). [Article]. *Journal of the Association for Information Science and Technology, 65*(4), 742-750. doi: 10.1002/asi.23075

Ye, F. Y., Bornmann, L., & Leydesdorff, L. (2017). h-based *I3*-type multivariate vectors: multidimensional indicators of publication and citation scores. *COLLNET Journal of Scientometrics and Information Management, 11*(1), 153-171.

Zsindely, S., Schubert, A., & Braun, T. (1982). Editorial Gatekeeping Patterns in International Science Journals -- A New Science Indicator. *Scientometrics, 4*(1), 57-68.